# Data Group Anonymity: General Approach

Oleg Chertov

Applied Mathematics Department
NTUU "Kyiv Polytechnic Institute"
Kyiv, Ukraine
chertov@i.ua

Dan Tavrov

Applied Mathematics Department
NTUU "Kyiv Polytechnic Institute"
Kyiv, Ukraine
dan.tavrov@i.ua

*Abstract*—In the recent time, the problem of protecting privacy in statistical data before they are published has become a pressing one. Many reliable studies have been accomplished, and loads of solutions have been proposed.

Though, all these researches take into consideration only the problem of protecting individual privacy, i.e., privacy of a single person, household, etc. In our previous articles, we addressed a completely new type of anonymity problems. We introduced a novel kind of anonymity to achieve in statistical data and called it group anonymity.

In this paper, we aim at summarizing and generalizing our previous results, propose a complete mathematical description of how to provide group anonymity, and illustrate it with a couple of real-life examples.

## I. INTRODUCTION

Throughout mankind's history, people always collected large amounts of demographical data. Though, until the very recent time, such huge data sets used to be inaccessible for publicity. And what is more, even if some potential intruder got an access to such paper-written data, it would be way too hard for him to analyze them properly!

But, as information technologies develop more, a greater number of specialists (to wide extent) gain access to large statistical datasets to perform various kinds of analysis. For that matter, different data mining systems help to determine data features, patterns, and properties.

As a matter of fact, in today world, in many cases population census datasets (usually referred to as *microfiles*) contain this or that kind of sensitive information about respondents. Disclosing such information can violate a person's privacy, so convenient precautions should be taken beforehand.

For many years now, mostly every paper in major of providing data anonymity deals with a problem of protecting an individual's privacy within a statistical dataset. As opposed to it, we have previously introduced a totally new kind of anonymity in a microfile which we called *group anonymity*. In this paper, we aim at gathering and systematizing all our works published in the previous years. Also, we would like to generalize our previous approaches and propose an integrated survey of group anonymity problem.

## II. RELATED WORK

### A. Individual Anonymity

We understand by *individual data anonymity* a property of information about an individual to be unidentifiable within a dataset.

There exist two basic ways to protect information about a single person. The first one is actually protecting the data in its formal sense, using data encryption, or simply restricting access to them. Of course, this technique is of no interest to statistics and affiliated fields.

The other approach lies in modifying initial microfile data such way that it is still useful for the majority of statistical researches, but is protected enough to conceal any sensitive information about a particular respondent. Methods and algorithms for achieving this are commonly known as *privacy preserving data publishing (PPDP)* techniques. The Free Haven Project [1] provides a very well prepared anonymity bibliography concerning these topics.

In [2], the authors investigated all main methods used in PPDP, and introduced a systematic view of them. In this subsection, we will only slightly characterize the most popular PPDP methods of providing individual data anonymity. These methods are also widely known as *statistical disclosure control (SDC)* techniques.

All SDC methods fall into two categories. They can be either *perturbative* or *non-perturbative*. The first ones achieve data anonymity by introducing some data distortion, whereas the other ones anonymize the data without altering them.

Possibly the simplest perturbative proposition is to add some noise to initial dataset [3]. This is called *data randomization*. If this noise is independent of the values in a microfile, and is relatively small, then it is possible to perform statistical analysis which yields rather close results compared to those ones obtained using initial dataset. Though, this solution is not quite efficient. As it was shown in [4], if there are other sources available aside from our microfile with intersecting information, it will be very possible to violate privacy.

Another option is to reach data *k-anonymity*. The core of this approach is to somehow ensure that all combinations of microfile attribute values are associated with at least *k* respondents. This result can be obtained using various methods [5, 6].

Yet another technique is to *swap* confidential microfile attribute values between different individuals [7].

Non-perturbative SDC methods are mainly represented by *data recoding* (data enlargement) and *data suppression* (removing the data from the original microfile) [6].

In previous years, novel methods evolved, e.g., matrix decomposition [8], or factorization [9]. But, all of them aim at preserving individual privacy only.

### B. Group Anonymity

Despite the fact that PPDP field is developing rather rapidly, there exists another, completely different privacy issue which hasn't been studied well enough yet. Speaking more precisely, it is another kind of anonymity to be achieved in a microfile.

We called this kind of anonymity *group anonymity*. The formal definition will be given further on in this paper, but in a way this kind of anonymity aims at protecting such data features and patterns which cannot be determined by analyzing standalone respondents.

The problem of providing group anonymity was initially addressed in [10]. Though, there has not been proposed any feasible solution to it then.

In [11, 12], we presented a rather effective method for solving some particular group anonymity tasks. We showed its main features, and discussed several real-life practical examples.

The most complete survey of group anonymity tasks and their solutions as of time this paper is being written is [13]. There, we tried to gather up all existing works of ours in one place, and also added new examples that reflect interesting peculiarities of our method. Still, [13] lacks a systematized view and reminds more of a collection of separate articles rather than of an integrated study.

That is why in this paper we set a task of embedding all known approaches to solving group anonymity problem into complete and consistent group anonymity theory.

## III. Formal Definitions

To start with, let us propose some necessary definitions.

*Definition 1.* By *microdata* we will understand various data about respondents (which might equally be persons, households, enterprises, and so on).

*Definition 2.* Respectively, we will consider *a microfile* to be microdata reduced to one file of attributive records concerning each single respondent.

A microfile can be without any complications presented in a matrix form. In such a matrix $\mathbf{M}$, each row corresponds to a particular respondent, and each column stands for a specific attribute. The matrix itself is shown in Table I.

| | | **Attributes** | | | |
|---|---|---|---|---|---|
| | | $u_1$ | $u_2$ | ... | $u_\eta$ |
| **Respondents** | $r_1$ | $\omega_{11}$ | $\omega_{12}$ | ... | $\omega_{1\eta}$ |
| | $r_2$ | $\omega_{21}$ | $\omega_{22}$ | ... | $\omega_{2\eta}$ |
| | ... | ... | ... | ... | ... |
| | $r_\mu$ | $\omega_{\mu 1}$ | $\omega_{\mu 2}$ | ... | $\omega_{\mu\eta}$ |

In such a matrix, we can define different classes of attributes.

*Definition 3. An identifier* is a microfile attribute which unambiguously determines a certain respondent in a microfile.

From a privacy protection point of view, identifiers are the most security-intensive attributes. The only possible way to prevent privacy violation is to completely eliminate them from a microfile. That is why, we will further on presume that a microfile is always *de-personalized*, i.e., it does not contain any identifiers.

In terms of group anonymity problem, we need to define such attributes whose distribution is of a big privacy concern and has to be thoroughly considered.

*Definition 4.* We will call an element $s_k^{(v)} \in S_v$, $k = \overline{1, l_v}$, $l_v \le \mu$ , where $S_v$ is a subset of a Cartesian product $u_{v_1} \times u_{v_2} \times ... \times u_{v_t}$ (see Table I), *a vital value combination*. Each element of $s_k^{(v)}$ is called *a vital value*. Each $u_{v_j}$, $j = \overline{1, t}$ is called *a vital attribute*.

In other words, vital attributes reflect characteristic properties needed to define a subset of respondents to be protected.

But, it is always convenient to present multidimensional data in a one-dimensional form to simplify its modification. To be able to accomplish that, we have to define yet another class of attributes.

*Definition 5.* We will call en element $s_k^{(p)} \in S_p$, $k = \overline{1, l_p}$, $l_p \le \mu$ , where $S_p$ is a subset of microfile data elements corresponding to the $p^{\text{th}}$ attribute, *a parameter value*. The attribute itself is called *a parameter attribute*.

Parameter values are usually used to somehow arrange microfile data in a particular order. In most cases, resultant data representation contains some sensitive information which is highly recommended to be protected. (We will delve into this problem in the next section.)

*Definition 6. A group* $G(V, P)$ is a set of attributes consisting of several vital attributes $V = \{V_1, V_2, ..., V_l\}$ and a parameter attribute $P$, $P \ne V_j$, $j = 1, ..., l$ .

Now, we can formally define a group anonymity task.

*Group Anonymity Definition.* The task of *providing data group anonymity* lies in modifying initial dataset for each group $G_i(V_i, P_i)$, $i = 1, ..., k$ such way that sensitive data features become totally confided.

In the next section, we will propose a generic algorithm for providing group anonymity in some most common practical cases.

## IV. GENERAL APPROACH TO PROVIDING GROUP ANONYMITY

According to the Group Anonymity Definition, initial dataset **M** should be perturbed separately for each group to ensure protecting specific features for each of them.

Before performing any data modifications, it is always necessary to preliminarily define what features of a particular group need to be hidden. So, we need to somehow transform initial matrix into another representation useful for such identification. Besides, this representation should also provide more explicit view of how to modify the microfile to achieve needed group features.

All this leads to the following definitions.

*Definition 7.* We will understand by *a goal representation* $\Omega$ (**M**, $G$) of a dataset **M** with respect to a group $G$ such a dataset (which could be of any dimension) that represents particular features of a group within initial microfile in a way appropriate for providing group anonymity.

We will discuss different forms of goal representations a bit later on in this section.

Having obtained goal representation of a microfile dataset, it is almost always possible to *modify* it such way that security-intensive peculiarities of a dataset become concealed. In this case, it is said we obtain *a modified goal representation* $\Omega'$ (**M**, $G$) of initial dataset **M**.

After that, we need to somehow map our modified goal representation to initial dataset resulting in *a modified microdata* **M**\*. Of course, it is not necessary that such data modifications lead to any feasible solution. But, as we will discuss it in the next subsections, if to pick specific mappings and data representations, it is possible to provide group anonymity in any microfile.

So, a generic scheme of providing group anonymity is as follows:

*1) Construct a (depersonalized) microfile* **M** *representing statistical data to be processed.*

*2) Define one or several groups* $G_i(V_i, P_i)$, $i = 1, ..., k$ *representing categories of respondents to be protected.*

*3) For each i from 1 to k:*

*a) Choosing data representation*: Pick a data representation $\Omega_i$ (**M**, $G_i$) for a group $G_i(V_i, P_i)$.

*b) Performing data mapping*: Define a mapping function $\Upsilon$ : **M** $\rightarrow \Omega_i$ (**M**, $G_i$) (called *goal mapping function*) and obtain needed goal representation of a dataset.

*c) Performing goal representation's modification*: Define a functional $\Xi$ : $\Omega_i$ (**M**, $G_i$) $\rightarrow \Omega'_i$ (**M**, $G_i$) (also called *modifying functional*) and obtain a modified goal representation.

*d) Obtaining the modified microfile*. Define an *inverse goal mapping function* $\Upsilon^{-1}$: $\Omega'_i$ (**M**, $G_i$) $\rightarrow$ **M**\* and obtain a modified microfile.

*4) Prepare the modified microfile for publishing.*

Now, let us discuss some of these algorithm steps a bit in detail.

### A. Different Ways to Construct a Goal Representation

In general, each particular case demands developing certain data representation models to suit the stated requirements the best way. Although, there are loads of real-life examples where some common models might be applied with a reasonable effect.

In our previous works, we drew a particular attention to one special data goal representation, namely, *a goal signal*. The goal signal is a one-dimensional numerical array $\theta = (\theta_1, \theta_2, ..., \theta_m)$ representing statistical features of a group. It can consist of values obtained in different ways, but we will defer this discussion for some paragraphs.

In the meantime, let us try to figure out what particular features of a goal signal might turn out to be security-intensive. To be able to do that, we need to consider its graphical representation which we will call *a goal chart*. In [13], we summarized the most important goal chart features and proposed some approaches to modifying them. In order not to repeat ourselves, we will only outline the most important ones:

*1) Extremums.* In most cases, it is the most sensitive information; we need to transit such extremums from one signal position to another (or, which is also completely convenient, create some new extremums, so that initial ones just "dissolve").

*2) Statistical features.* Such features as signal mean value and standard deviation might be of a big importance, unless a corresponding parameter attribute is nominal (it will become clear why in a short time).

*3) Frequency spectrum.* This feature might be rather interesting if a goal signal contains some parts repeated cyclically.

Coming from a particular aim to be achieved, one can choose the most suitable modifying functional $\Xi$ to redistribute the goal signal.

Let us understand how a goal signal can be constructed in some widely spread real-life group anonymity problems.

In many cases, we can count up all the respondents in a group with a certain pair of vital value combination and a parameter value, and arrange them in any order proper for a parameter attribute. For instance, if parameter values stand for a person's age, and vital value combinations reflect his or her yearly income, then we will obtain a goal signal representing quantities of people with a certain income distributed by their age. In some situations, this distribution could lead to unveiling some restricted information, so, a group anonymity problem would evidently arise.

Such a goal signal is called *a quantity signal* $q = (q_1, q_2, ..., q_m)$. It provides a quantitative statistical distribution of group members from initial microfile.

Though, as it was shown in [12], sometimes absolute quantities do not reflect real situations, because they do not take into account all the information given in a microfile. A much better solution for such cases is to build up *a concentration signal*:

$$c = (c_1, c_2, ..., c_m) \equiv \left( \frac{q_1}{\rho_1}, \frac{q_2}{\rho_2}, ..., \frac{q_m}{\rho_m} \right). \qquad (1)$$

In (1), $\rho_i$, $i = 1, ..., m$ stand for the quantities of respondents in a microfile from a group defined by a superset for our vital value combinations. This can be explained on a simple example. Information about people with AIDS distributed by regions of a state can be valid only if it is represented in a relative form. In this case, $q_i$ would stand for a number of ill people in the $i^{th}$ region, whereas $\rho_i$ could possibly stand for the whole number of people in the $i^{th}$ region.

And yet another form of a goal signal comes to light when processing comparative data. A representative example is as follows: if we know concentration signals built separately for young males of military age and young females of the same age, then, maximums in their difference might point at some restricted military bases.

In such cases, we deal with two concentration signals $c^{(1)} = (c_1^{(1)}, c_2^{(1)}, ..., c_m^{(1)})$ (also called *a main concentration signal*) and $c^{(2)} = (c_1^{(2)}, c_2^{(2)}, ..., c_m^{(2)})$ (*a subordinate concentration signal*). Then, the goal signal takes a form of *a concentration difference signal* $\delta = (c_1^{(1)} - c_1^{(2)}, c_2^{(1)} - c_2^{(2)}, ..., c_m^{(1)} - c_m^{(2)})$.

In the next subsection, we will address the problem of picking a suitable modifying functional, and also consider one of its possible forms already successfully applied in our previous papers.

### B. Picking Appropriate Modifying Functional

Once again, there can be created way too many unlike modifying functionals, each of them taking into consideration these or those requirements set by a concrete group anonymity problem definition. In this subsection, we will look a bit in detail at two such functionals.

So, let us pay attention to the first goal chart feature stated previously, which is in most cases the feature we would like to protect. Let us discuss the problem of altering extremums in an initial goal chart.

In general, we might perform this operation quite arbitrarily. The particular scheme of such extremums redistribution would generally depend on the quantity signal nature, sense of parameter values, and correct data interpreting. But, as things usually happen in statistics, we might as well want to guarantee that data utility wouldn't reduce much. By *data utility preserving* we will understand the situation when the modified goal signal yields similar, or even the same, results when performing particular statistical (but not exclusively) analysis.

Obviously, altering the goal signal completely off-hand without any additional precautions taken wouldn't be very convenient from the data utility preserving point of view. Hopefully, there exist two quite dissimilar, thought powerful techniques for preserving some goal chart features.

The first one was proposed in [14]. Its main idea is to *normalize* the output signal using such transformation that both mean value and standard deviation of a signal remain stable. Surely, this is not ideal utility preserving. But, the signal obtained this way at least yields the same results when performing basic statistical analysis. So, the formula goes as follows:

$$\theta^* = (\theta + \frac{\sigma^*}{\sigma} \cdot \varepsilon - \varepsilon^*) \cdot \frac{\sigma}{\sigma^*}. \qquad (2)$$

In (2), $\varepsilon = \frac{1}{m} \sum_{i=1}^{m} \theta_i$, $\varepsilon^* = \frac{1}{m} \sum_{i=1}^{m} \theta_i^*$, $\sigma = \sqrt{\frac{\sum_{i=1}^{m}(\theta_i - \varepsilon)^2}{m-1}}$, $\sigma^* = \sqrt{\frac{\sum_{i=1}^{m}(\theta_i^* - \varepsilon^*)^2}{m-1}}$.

The second method of modifying the signal was initially proposed in [11], and was later on developed in [12, 13]. Its basic idea lies in applying *wavelet transform* to perturbing the signal, with some slight restrictions necessary for preserving data utility:

$$\theta(t) = \sum_i a_{k,i} \cdot \varphi_{k,i}(t) + \sum_{j=k}^{1} \sum_i d_{j,i} \cdot \psi_{j,i}(t). \qquad (3)$$

In (3), $\varphi_{k,i}$ stands for shifted and sampled *scaling functions*, and $\psi_{j,i}$ represents shifted and sampled *wavelet functions*. As we showed in our previous researches, we can gain group anonymity by modifying *approximation coefficients* $a_{k,i}$. At the same time, if we don't modify *detail coefficients* $d_{j,i}$ we can preserve signal's frequency characteristics necessary for different kinds of statistical analysis.

More than that, we can always preserve the signal's mean value without any influence on its extremums:

$$\theta_{fin}^{*} = \theta_{mod}^{*} \cdot \left( \sum_{i=1}^{m} \theta_i \Big/ \sum_{i=1}^{m} \theta_{mod\ i}^{*} \right). \qquad (4)$$

In the next section, we will study several real-life practical examples, and will try to provide group anonymity for appropriate datasets. Until then, we won't delve deeper into wavelet transforms theory.

### C. The Problem of Minimum Distortion when Applying Inverse Goal Mapping Function

Having obtained modified goal signal $\theta_{fin}^{*}$, we have no other option but to modify our initial dataset $\mathbf{M}$, so that its contents correspond to $\theta_{fin}^{*}$.

It is obvious that, since group anonymity has been provided with respect to only a single respondent group, modifying the dataset $\mathbf{M}$ almost inevitably will lead to introducing some level of data distortion to it. In this subsection, we will try to minimize such distortion by picking sufficient inverse goal mapping functions.

At first, we need some more definitions.

*Definition 8.* We will call microfile $\mathbf{M}$ attributes *influential ones* if their distribution plays a great role for researchers.

Obviously, vital attributes are influential by definition.

Keeping in mind this definition, let us think over a particular procedure of mapping the modified goal signal $\theta_{fin}^{*}$ to a modified microfile $\mathbf{M}^{*}$. The most adequate solution, in our opinion, implies swapping parameter values between pairs of somewhat close respondents. We might interpret this operation as "transiting" respondents between two different groups (which is in fact the case).

But, an evident problem arises. We need to know how to define whether two respondents are "close" or not. This could be done if to measure such closeness using *influential metric* [13]:

$$InfM(r, r^{*}) = \sum_{p=1}^{n_{ord}} \zeta_p \left( \frac{r(I_p) - r^{*}(I_p)}{r(I_p) + r^{*}(I_p)} \right)^2 +$$
$$+ \sum_{k=1}^{n_{nom}} \gamma_k \left( \chi \big( r(J_k), r^{*}(J_k) \big) \right)^2. \qquad (5)$$

In (5), Here, $I_p$ stands for the $p^{th}$ ordinal influential attribute (making a total of $n_{ord}$). Respectively, $J_k$ stands for the $k^{th}$ nominal influential attribute (making a total of $n_{nom}$). Functional $r(\cdot)$ stands for a record's $r$ specified attribute value. Operator $\chi(v_1, v_2)$ is equal to $\chi_1$ if values $v_1$ and $v_2$ represent one category, and $\chi_2$, if it is not so. Coefficients $\zeta_p$ and $\gamma_k$ should be taken coming from importance of a certain attribute (for those ones not to be changed at all they ought to be as big as possible, and for those that are not important they could be zero).

With the help of this metric, it is not too hard to outline the generic strategy of performing inverse data mapping. One needs to search for every pair of respondents yielding minimum influential metric value, and swap corresponding parameter values. This procedure should be carried out until the modified goal signal $\theta_{fin}^{*}$ is completely mapped to $\mathbf{M}^{*}$.

This strategy seems to be NP-hard, so, the problem of developing more computationally effective inverse goal mapping functions remains open.

## V. SOME PRACTICAL EXAMPLES OF PROVIDING GROUP ANONYMITY

In this subsection, we will discuss two practical examples built upon real data to show the proposed group anonymity providing technique in action.

According to the scheme introduced in Section IV, the first thing to accomplish is to compile a microfile representing the data we would like to work with. For both of our examples, we decided to take 5-Percent Public Use Microdata Sample Files provided by the U.S. Census Bureau [15] concerning the 2000 U.S. census of population and housing microfile data. But, since this dataset is huge, we decided to limit ourselves with analyzing the data on the state of California only.

The next step (once again, we will carry it out the same way for both examples) is to define group(s) to be protected. In this paper, we will follow [11], i.e. we will set a task of protecting military personnel distribution by the places they work at. Such a task has a very important practical meaning. The thing is that extremums in goal signals (both quantity and concentration ones) with a very high probability mark out the sites of military cantonments. In some cases, these cantonments aren't likely to become widely known (especially to some potential adversaries).

So, to complete the second step of our algorithm, we take "Military service" attribute as a vital one. This is a categorical attribute, with integer values ranging from 0 to 4. For our task definition, we decided to take one vital value, namely, "1" which stands for "Active duty".

But, we also need to pick an appropriate parameter attribute. Since we aim at redistributing military servicemen by different territories, we took "Place of Work Super-PUMA" as a parameter attribute. The values of this categorical attribute represent codes for Californian statistical areas. In order to simplify our problem a bit, we narrowed the set of this attribute's values down to the following ones: 06010, 06020, 06030, 06040, 06060, 06070, 06080, 06090, 06130, 06170, 06200, 06220, 06230, 06409, 06600, and 06700. All these area codes correspond to border, island, and coastal statistical areas.

From this point, we need to make a decision about the goal representation of our microdata. To show peculiarities of different kinds of such representations, we will discuss at least two of them in this section. The first one would be the quantity signal, and the other one would be its concentration analogue.

## A. Quantity Group Anonymity Problem

So, having all necessary attributes defined, it is not too hard to count up all the military men in each statistical area, and gather them up in a numerical array sorted in an ascending order by parameter values. In our case, this quantity signal looks as follows:

$$q = (19, 12, 153, 71, 13, 79, 7, 33, 16, 270, 812, 135, 241, 14, 60, 4337).$$

The graphical representation of this signal is presented in Fig. 1a.

As we can clearly see, there is a very huge extremum at the last signal position. So, we need to somehow eliminate it, but simultaneously preserve important signal features. In this example, we will use wavelet transforms to transit extremums to another region, so, according to the previous section, we will be able to preserve high-frequency signal spectrum.

As it was shown in [11], we need to change signal approximation coefficients in order to modify its distribution. To obtain approximation coefficients of any signal, we need to *decompose* it using appropriate *wavelet filters* (both high- and low-frequency ones). We won't explain in details here how to perform all the wavelet transform steps (refer to [12] for details), though, we will consider only those steps which are necessary for completing our task.

So, to decompose the quantity signal $q$ by two levels using Daubechies second-order low-pass wavelet decomposition filter $l \equiv \left( \dfrac{1-\sqrt{3}}{4\sqrt{2}}, \dfrac{3-\sqrt{3}}{4\sqrt{2}}, \dfrac{3+\sqrt{3}}{4\sqrt{2}}, \dfrac{1+\sqrt{3}}{4\sqrt{2}} \right)$, we need to perform the following operations:

$$a_2 = (q *_{\downarrow 2} l) *_{\downarrow 2} l = (2272.128, 136.352, 158.422, 569.098).$$

By $*_{\downarrow 2}$ we denote the operation of convolution of two vectors followed by dyadic downsampling of the output. Also, we present the numerical values with three decimal numbers only due to the limited space of this paper.

By analogue, we can use the flipped version of $l$ (which would be a high-pass wavelet decomposition filter) denoted by $h = \left( \dfrac{1+\sqrt{3}}{4\sqrt{2}}, \dfrac{3+\sqrt{3}}{4\sqrt{2}}, \dfrac{3-\sqrt{3}}{4\sqrt{2}}, \dfrac{1-\sqrt{3}}{4\sqrt{2}} \right)$ to obtain detail coefficients at level 2:

$$d_2 = (q *_{\downarrow 2} l) *_{\downarrow 2} h = (-508.185, 15.587, 546.921, -315.680).$$

According to the wavelet theory, every numerical array can be presented as the sum of its low-frequency component (at the last decomposition level) and a set of several high-frequency ones at each decomposition level (called *approximation* and *details* respectively). In general, the signal approximation and details can be obtained the following way (we will also substitute the values from our example):

$$A_2 = (a_2 *_{\uparrow 2} l) *_{\uparrow 2} l = (1369.821, 687.286, 244.677, 41.992, -224.980, 11.373, 112.860, 79.481, 82.240, 175.643, 244.757, 289.584, 340.918, 693.698, 965.706, 1156.942);$$

$$D_1 + D_2 = d_1 *_{\uparrow 2} h + (d_2 *_{\uparrow 2} h) *_{\uparrow 2} l = (-1350.821, -675.286, -91.677, 29.008, 237.980, 67.627, -105.860, -46.481, -66.240, 94.357, 567.243, -154.584, -99.918, -679.698, -905.706, 3180.058).$$

To provide group anonymity (or, redistribute signal extremums, which is the same), we need to replace $A_2$ with another approximation, such that the resultant signal (obtained when being summed up with our details $D_1 + D_2$) becomes different. Moreover, the only values we can try to alter are approximation coefficients.

So, in general, we need to solve a corresponding optimization problem. Knowing the dependence between $A_2$ and $a_2$ (which is pretty easy to obtain in our model example), we can set appropriate constraints, and obtain a solution $\hat{a}_2$ which completely meets our requirements.

For instance, we can set the following constraints:

$$\begin{cases} 0.637 \cdot \hat{a}_2(1) - 0.137 \cdot \hat{a}_2(4) \le 1369.821; \\ 0.296 \cdot \hat{a}_2(1) + 0.233 \cdot \hat{a}_2(2) - 0.029 \cdot \hat{a}_2(4) \le 687.286; \\ 0.079 \cdot \hat{a}_2(1) + 0.404 \cdot \hat{a}_2(2) + 0.017 \cdot \hat{a}_2(4) \le 244.677; \\ -0.137 \cdot \hat{a}_2(1) + 0.637 \cdot \hat{a}_2(2) \ge -224.980; \\ -0.029 \cdot \hat{a}_2(1) + 0.296 \cdot \hat{a}_2(2) + 0.233 \cdot \hat{a}_2(3) \ge 11.373; \\ 0.017 \cdot \hat{a}_2(1) + 0.079 \cdot \hat{a}_2(2) + 0.404 \cdot \hat{a}_2(3) \ge 112.860; \\ -0.012 \cdot \hat{a}_2(2) + 0.512 \cdot \hat{a}_2(3) \ge 79.481; \\ -0.137 \cdot \hat{a}_2(2) + 0.637 \cdot \hat{a}_2(3) \ge 82.240; \\ -0.029 \cdot \hat{a}_2(2) + 0.296 \cdot \hat{a}_2(3) + 0.233 \cdot \hat{a}_2(4) \ge 175.643; \\ 0.233 \cdot \hat{a}_2(1) - 0.029 \cdot \hat{a}_2(3) + 0.296 \cdot \hat{a}_2(4) \ge 693.698; \\ 0.404 \cdot \hat{a}_2(1) + 0.017 \cdot \hat{a}_2(3) + 0.079 \cdot \hat{a}_2(4) \le 965.706; \\ 0.512 \cdot \hat{a}_2(1) - 0.012 \cdot \hat{a}_2(4) \le 1156.942. \end{cases}$$

The solution might be as follows: $\hat{a}_2 = (0, 379.097, 31805.084, 5464.854)$.

Now, let us obtain our new approximation $\hat{A}_2$, and a new quantity signal $\hat{q}$:

$$\hat{A}_2 = (\hat{a}_2 *_{\uparrow 2} l) *_{\uparrow 2} l = (-750.103, -70.090, 244.677, 194.196, 241.583, 345.372, 434.049, 507.612, 585.225, 1559.452, 2293.431, 2787.164, 3345.271, 1587.242, 449.819, -66.997);$$

$$\hat{q} = \hat{A}_2 + D_1 + D_2 = (-2100.924, -745.376, 153.000, 223.204, 479.563, 413.000, 328.189, 461.131, 518.985, 1653.809, 2860.674, 2632.580, 3245.352, 907.543, -455.887, 3113.061).$$

Two main problems almost always arise at this stage. As we can see, there are some negative elements in the modified goal signal. This is completely awkward. A very simple though quite adequate way to overcome this backfire is to add a reasonably big number (2150 in our case) to all signal elements. Obviously, the mean value of the signal will change. After all, these two issues can be solved using the following formula: $q_{mod}^{*} = (\hat{q} + 2150) \cdot \left( \sum_{i=1}^{16} q_i \right) / \left( \sum_{i=1}^{16} (\hat{q}_i + 2150) \right)$.

If to round $q_{mod}^{*}$ (since quantities have to be integers), we obtain the modified goal signal as follows:

$q_{fin}^{*} = $ (6, 183, 300, 310, 343, 334, 323, 341, 348, 496, 654, 624, 704, 399, 221, 686).

The graphical representation is available in Fig. 1b.

As we can see, the group anonymity problem at this point has been completely solved: all initial extremums persisted, and some new ones emerged.

The last step of our algorithm (i.e., obtaining new microfile $\mathbf{M}^{*}$) cannot be shown in this paper due to evident space limitations.

### B. Concentration Group Anonymity Problem

Now, let us take the same dataset we processed before. But, this time we will pick another goal mapping function. We will try to build up a concentration signal.

According to (1), what we need to do first is to define what $\rho_l$ to choose. In our opinion, the whole quantity of males 18 to 70 years of age would suffice.

By completing necessary arithmetic operations, we finally obtain the concentration signal:

$c = $ (0.004, 0.002, 0.033, 0.009, 0.002, 0.012, 0.002, 0.007, 0.001, 0.035, 0.058, 0.017, 0.030, 0.003, 0.004, 0.128).

The graphical representation can be found in Fig. 2a.

Let us perform all the operations we've accomplished earlier, without any additional explanations (we will reuse notations from the previous subsection):

$a_2 = (c *_{\downarrow 2} l) *_{\downarrow 2} l = $ (0.073, 0.023, 0.018, 0.059);

$d_2 = (c *_{\downarrow 2} l) *_{\downarrow 2} h = $ (0.003, −0.001, 0.036, −0.018);

$A_2 = (a_2 *_{\uparrow 2} l) *_{\uparrow 2} l = $ (0.038, 0.025, 0.016, 0.011, 0.004, 0.009, 0.010, 0.009, 0.008, 0.019, 0.026, 0.030, 0.035, 0.034, 0.034, 0.037);

$D_1 + D_2 = d_1 *_{\uparrow 2} h + (d_2 *_{\uparrow 2} h) *_{\uparrow 2} l = $ (−0.034, −0.023, 0.017, −0.002, −0.002, 0.003, −0.009, −0.002, −0.007, 0.016, 0.032, −0.013, −0.005, −0.031, −0.030, 0.091).

The constraints for this example might look the following way:

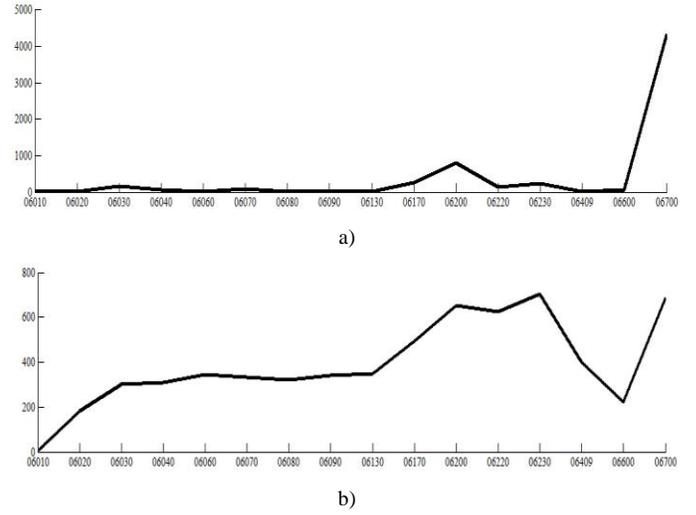

Figure 1.   Initial (a) and modified (b) quantity signals.

$$\begin{cases} 0.637 \cdot \hat{a}_2(1) - 0.137 \cdot \hat{a}_2(4) \leq 0.038; \\ 0.296 \cdot \hat{a}_2(1) + 0.233 \cdot \hat{a}_2(2) - 0.029 \cdot \hat{a}_2(4) \leq 0.025; \\ 0.079 \cdot \hat{a}_2(1) + 0.404 \cdot \hat{a}_2(2) + 0.017 \cdot \hat{a}_2(4) \leq 0.016; \\ -0.012 \cdot \hat{a}_2(1) + 0.512 \cdot \hat{a}_2(2) \leq 0.011; \\ -0.137 \cdot \hat{a}_2(1) + 0.637 \cdot \hat{a}_2(2) \geq 0.005; \\ -0.029 \cdot \hat{a}_2(1) + 0.296 \cdot \hat{a}_2(2) + 0.233 \cdot \hat{a}_2(3) \geq 0.009; \\ 0.017 \cdot \hat{a}_2(1) + 0.079 \cdot \hat{a}_2(2) + 0.404 \cdot \hat{a}_2(3) \geq 0.010; \\ -0.012 \cdot \hat{a}_2(2) + 0.512 \cdot \hat{a}_2(3) \geq 0.009; \\ -0.137 \cdot \hat{a}_2(2) + 0.637 \cdot \hat{a}_2(3) \geq 0.009; \\ -0.029 \cdot \hat{a}_2(2) + 0.296 \cdot \hat{a}_2(3) + 0.233 \cdot \hat{a}_2(4) \geq 0.019; \\ 0.233 \cdot \hat{a}_2(1) - 0.029 \cdot \hat{a}_2(3) + 0.296 \cdot \hat{a}_2(4) \leq 0.034; \\ 0.404 \cdot \hat{a}_2(1) + 0.017 \cdot \hat{a}_2(3) + 0.079 \cdot \hat{a}_2(4) \leq 0.034; \\ 0.512 \cdot \hat{a}_2(1) - 0.012 \cdot \hat{a}_2(4) \leq 0.037. \end{cases}$$

One possible solution to this system is as follows: $\bar{a}_2 = $ = (0, 0.002, 0.147, 0.025).

We can obtain new approximation and concentration signal:

$\hat{A}_2 = (\bar{a}_2 *_{\uparrow 2} l) *_{\uparrow 2} l = $ (−0.003, −0.000, 0.001, 0.001, 0.001, 0.035, 0.059, 0.075, 0.093, 0.049, 0.022, 0.011, −0.004, 0.003, 0.005, 0.000);

$\hat{c} = \hat{A}_2 + D_1 + D_2 = $ (−0.037, −0.023, 0.018, −0.001, −0.002, 0.038, 0.051, 0.073, 0.086, 0.066, 0.054, −0.002, −0.009, −0.028, −0.026, 0.092).

Once again, we need to make our signal non-negative, and fix its mean value. But, it is obvious that the corresponding quantity signal $q_{mod}^{*}$ will also have a different mean value. Therefore, fixing the mean value can be done in "the quantity domain" (which we won't present here).

Nevertheless, it is possible to make the signal non-negative after all:

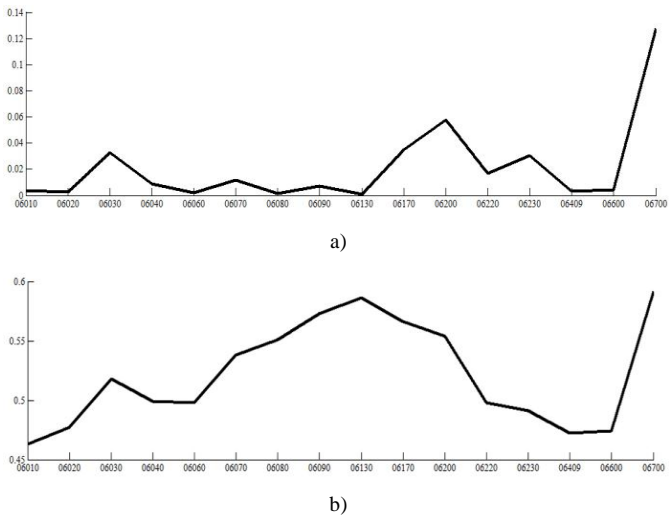

a)

b)

Figure 2. Initial (a) and modified (b) concentration signals.

$$c^*_{mod} = \hat{c} + 0.5 = (0.463, 0.477, 0.518, 0.499, 0.498, 0.538,$$
$$0.551, 0.573, 0.586, 0.566, 0.554, 0.498, 0.491, 0.472, 0.474,$$
$$0.592).$$

The graphical representation can be found in Fig. 2b. Once again, the group anonymity has been achieved.

Once again, the last step to complete is to construct the modified $\mathbf{M}^*$, which we will omit in this paper.

## VI. SUMMARY

In this paper, it is the first time that group anonymity problem has been thoroughly analyzed and formalized. We presented a generic mathematical model for group anonymity in microfiles, outlined the scheme for providing it in practice, and showed several real-life examples.

As we think, there still remain some unresolved issues, some of them are as follows:

*1) Choosing data representation:* There are still many more ways to pick convenient goal representation of initial data not covered in this paper. They might depend on some problem task definition peculiarities.

*2) Performing goal representation's modification:* It is obvious that the method discussed in Section V is not an exclusive one. There could be as well proposed other sufficient techniques to perform data modifications. For instance, choosing different wavelet bases could lead to yielding different outputs.

*3) Obtaining the modified microfile:* There has to be developed computationally effective heuristics to perform inverse goal mapping.